\documentstyle[12pt]{article}
\advance\voffset by 28 pt
\def\lept{l^+l^-}
\def\htau{\hat \tau}
\def\sin{\mbox{\rm sin}}
\def\cos{\mbox{\rm cos}}
\def\tan{\mbox{\rm tan}}
\def\cot{\mbox{\rm cot}}
\def\sec{\mbox{\rm sec}}

\def\t2{{\hat \theta \over 2}}
\def\annia{(\bar q_f(x_1) q_f(x_2) - \Delta \bar q_f(x_1)
\Delta q_f(x_2))}
\def\annib{(\Delta \bar q_f(x_1) q_f(x_2) - \bar q_f(x_1)
\Delta q_f(x_2))}
\def\gqa{G(x_1) q_f(x_2)}
\def\gqb{G(x_1) \Delta q_f(x_2)}
\def\gqc{\Delta G(x_1) q_f(x_2)}
\def\gqd{\Delta G(x_1) \Delta q_f(x_2)}
\begin{document}
\pagestyle{empty}
\noindent
{\flushright CERN-TH.6997/93\\}
\vspace{2cm}
\begin{center}
{\Large \bf \boldmath $\gamma^*$, $Z^*$ production in polarised $p$--$p$
scattering\\
as a probe of the proton spin structure}\\
\vspace{8mm}
{E. Leader}\\
\vspace{4mm}
{\em Department of Physics, Birkbeck College,\\
University of London, Malet Street, London WC1E 7HX, U.K.}\\
\vspace{8mm}
{K. Sridhar$^{*}$}\\
\vspace{4mm}
{\em Theory Division, CERN, CH-1211, Geneva 23, Switzerland.}\\
\end{center}

\vspace{1.5cm}
\begin{abstract}
We present the results of a detailed study of the large
transverse momentum Drell-Yan process, $pp \rightarrow ( \gamma^*,
\ Z^* ) X \rightarrow \lept X$ at collider energies, with either
one or both protons
polarised, allowing the study of single- and double-spin
asymmetries respectively. We show how these asymmetries
obtained from angular distributions of the leptons
in the $\gamma^*$ (or $Z^*$) rest frame, can be used to get
information on the polarised parton distributions. Numerical
results for the asymmetries and the cross-sections are presented,
and the sensitivity of the asymmetries to the initial parton
distributions indicates that these can be used as
effective probes of the spin structure of the proton.
\end{abstract}

\vspace{2cm}
\noindent
\vspace{1cm} $^{*)} $ sridhar@vxcern.cern.ch\\
CERN-TH.6997/93\\
September 1993\\

\vfill
\clearpage
\setcounter{page}{1}
\pagestyle{plain}
\section*{1. Introduction}
Over the last two decades, deep-inelastic scattering experiments
have given us important clues to the structure of the nucleon.
Many of the predictions of the QCD-improved parton model
have now been tested in these experiments, and we have a
reasonable understanding of the structure of the nucleon
in terms of its constituents~-- quarks and gluons. This
understanding, however, is far from complete. There are
several aspects of hadronic phenomenology that do not yield
to a complete description in terms of the QCD-improved parton
model. One striking example of this is the description of the
spin structure of the nucleon.

The limited amount of experimental information on the spin structure
of the nucleon that is available is from polarised deep-inelastic
scattering experiments~-- scattering of polarised leptons off
polarised nucleon targets. Over five years ago, the EMC collaboration
first published its results \cite{emc} obtained from polarised
muon-proton scattering experiments. For the polarised structure
function $g_1^p(x,Q^2)$, which the EMC experiment measured in the
range $0.015 \le x \le 0.7$, the first moment gives
\begin{equation}
\label{e1}
\Gamma_1^p \equiv \int_0^1 g_1^p(x,Q^2) dx = 0.128 \pm 0.032.
\end{equation}
This integral was obtained by assuming a smooth extrapolation based
on Regge behaviour for the unmeasured low-$x$ region $x \le 0.015$.
In the parton model, we have the relation
\begin{equation}
\label{e2}
g_1^p(x) = {1 \over 2} \sum_f e_f^2\Delta q^f(x).
\end{equation}
In the above equation, $\Delta q^f (x) = (q^f_+(x)-q^f_-(x))$, where
$q^f_{\pm}$ refer to densities of quarks with $\pm$ helicity in a
proton with helicity ${1\over 2}$, i.e. the
$\Delta q^f$'s are twice the contribution to the nucleon's spin of a
quark of flavour $f$. Data \cite{bourquin} from hyperon decays determine
two independent linear combinations of the first moments of
$\Delta q^f $'s. Using the information in Eq.~\ref{e1} with that
available from hyperon decays, it is possible to determine the
first moments of $\Delta q^f$'s ($i=u,d,s$) separately. This yields
for the integral of twice the sum of the spins of the quarks \cite{altar}
\begin{equation}
\label{e3}
\Delta \Sigma \equiv
\int \Delta \Sigma dx \equiv \int (\Delta u + \Delta d + \Delta s) dx
 = 0.12 \pm 0.17.
\end{equation}
This is the famous EMC result that the total spin carried
by the quarks is small and is actually compatible with zero.

More recently, data from the SMC muon-deuteron scattering
experiment \cite{smc} and from the SLAC muon-${}^3$He
scattering experiment \cite{slac} have become available.
Taken at face value, the SLAC results seem to be in disagreement
with the EMC and SMC results. However, a careful analysis
\cite{elkar} of all the uncertainties show that the values of
$\Delta\Sigma$ obtained from the three sets of data are not
incompatible with each other. A global average yields
\begin{equation}
\label{e4}
\Delta \Sigma = 0.22 \pm 0.10.
\end{equation}
This value is not very different from the value in Eq.~\ref{e3};
and the global average also gives a rather small value for
$\Delta\Sigma$. In the naive parton model, the theoretical problem
is to understand the smallness of this quantity which could arise
from a large, negative strange quark polarisation \cite {strange}.
However, in QCD, there is an alternative interpretation \cite{anomaly}
of the data on $\Gamma_1^p$. Via the axial anomaly, there arises a
gluonic contribution to $g_1^p(x)$ and to $\Gamma_1^p$ so that the
conclusion that $\Delta \Sigma$ is small can be avoided, if the gluonic
contribution is large enough, i.e. if the polarised gluon distribution
function is non-negligible.

It thus becomes important to have other independent information on
the polarised quark and gluon densities. In this paper, we study the
large transverse momentum Drell-Yan process, taking into account
$Z$-interference, as a probe of the gluon polarisation in the nucleon.
We use a range of polarised parton distributions, fitted to the data
from the EMC experiment. We study this process for
the energies ($\sqrt{s} \sim 500~$GeV) planned at the proposed polarised
$p$--$p$ experiment at the Relativistic Heavy Ion Collider (RHIC) in
Brookhaven. A detailed account of the proposal can be found in
Ref.~\cite{rhic}. At RHIC, experiments with both incident protons
polarised and those with only one of the incident protons polarised,
are envisaged. The process that we are interested in is $pp \rightarrow
l^+l^- + X$, where the lepton pair has large transverse momentum
and where either one or both initial protons are polarised,
allowing the study of single-spin and double-spin asymmetries,
respectively. For this process, we study the angular distributions of
the leptons, and demonstrate how this is sensitive to the
polarised parton distributions. At the high energies under
consideration, virtual photons and virtual $Z$'s contribute
to lepton pair production. The parity-non-conserving $Z$ vertex
is crucial to our considerations because this results in a
non-zero single-spin asymmetry.

Some of the results discussed in this paper, namely those relating
to the simplest asymmetries of the cross-sections, were published in an
earlier letter \cite{earlett}. In this paper, we discuss in detail
the theoretical framework for obtaining the angular distributions,
and present complete expressions for the additional asymmetries
associated with the multipole parameters.
Numerical results beyond that presented in Ref.~\cite{earlett},
are also included in the present paper. The remainder of this
paper is organised as follows~: in Sec.~2, we begin with a
discussion of the explanation of the EMC experiment in terms
of the axial anomaly and a large gluon polarisation and emphasise
the need to have independent experimental information on the
magnitude and the shape of the polarised gluon distributions.
After briefly discussing the proposals to measure the gluon
polarisation that exist in the literature, we discuss the
conditions under which a non-vanishing single-spin asymmetry
can be obtained. In Sec.~3, we discuss the process $pp \rightarrow
\lept + X$ and give the general framework for obtaining the
angular distributions of the leptons in the rest frame of the
vector boson ($\gamma^*$ or $Z^*$), and in Sec.~4 we present the
numerical results for the asymmetries and the cross-sections. We
state our conclusions in Sec.~5. In Appendix A, we list the complete
expressions for the density matrices and the multipole parameters, and
give the expressions for the decay parameters in Appendix B.

\section*{2. Gluon polarisation and the single-spin asymmetry}
We begin this section with a brief description of the anomaly
explanation \cite{anomaly} of polarised deep-inelastic scattering
experiments. In the naive parton model, the flavour-singlet
part of the first moment of the polarised structure functions
is given by
\begin{equation}
\label{e5}
\Gamma_1^p \vert_{singlet} = {\langle e^2 \rangle \over 2} \Delta \Sigma.
\end{equation}
However, in the QCD-improved parton model, this relation is no longer
true: there is an additional piece due to the interaction of the
photon with a gluon $via$ a quark loop. Naively, this contribution
would be expected to be small, as it involves an extra factor of
$\alpha_s$ as compared with the lowest-order photon-quark diagram.
But the contribution is not small because the longitudinally polarised
gluon distribution evolves in such a way as to compensate exactly
the logarithmic decrease with $Q^2$ of $\alpha_s$. For the first moment
of the singlet part of $g_1^p$, the gluon contribution is given by
the triangle axial anomaly, so the singlet part of the measured
polarised structure function includes a gluon contribution, which is
given by the following relation
\begin{equation}
\label{e6}
\Gamma_1^p \vert_{singlet} = {\langle e^2 \rangle \over 2} \Delta \Sigma
- {\alpha_s \over 2\pi} N_f \Delta g,
\end{equation}
where $N_f$ is the number of quark flavours that circulate in the
loop. The empirical smallness of $\Gamma_1^p \vert_{singlet}$ is
explained in this picture by postulating a cancellation between the
$\Delta \Sigma$ and the $\Delta g$ contributions; in order that this
cancellation occurs one requires a large gluon polarisation.

It is important to realise that the anomaly explanation of the EMC
experiment does not provide any estimate of $\Delta g$, neither do we
have any other theoretical reason to believe that it must be large.
An independent experimental determination of this quantity is
crucial, and will, in fact, help us understand the spin structure
of the nucleon. There have been several suggestions in the literature
\cite{pheno} on how the gluon polarisation may be experimentally
determined. Most of these are aimed at experiments where the beam
and the target are both longitudinally polarised~-- the asymmetries
that can be studied in this case are the double-spin asymmetries.
{}From the experimental point of view, however, it would be much simpler
if only one of the initial particles were polarised. There exist in
the literature few suggestions on how the single-spin asymmetries
thus measured may be used to determine the polarised parton densities.

The most direct way of measuring a single-spin asymmetry is to
measure the polarisation of a final-state particle, and the
measured asymmetry would then simply correlate the final polarisation
to the initial polarisation. The measurement of the polarisation of
a final-state photon in direct photon production \cite{taxil}
or of a final-state lepton in Drell-Yan dilepton production \cite{conto}
have, therefore, been suggested. However, it is not an easy task to
measure the polarisation of a final-state particle at high energies,
and, hence, it becomes necessary to look for other ways of obtaining
a finite single-spin asymmetry. One way is to measure at least two
final-state momenta and couple them through their cross-product to the
spin of the initial particle. The cross-product and the spin are both
axial vectors and yield a scalar correlation; this correlation, however,
is $T$-odd and can, therefore, be induced only through loop corrections.
A detailed study of one-loop corrections to polarised Drell-Yan has been
made \cite{carlitz}, in order to estimate the size of the asymmetry
arising from the imaginary part of the amplitudes. The only other
way of obtaining a non-vanishing single-spin asymmetry is to consider
parity-violating processes like $W$ production \cite{leader, weber}.
This asymmetry obtains because the initial spin couples with
the axial part of the $W$-fermion vertex. In this paper, we
consider another parity-violating process: large $p_T$ Drell-Yan
with $\gamma^*$--$Z^*$ interference, and study both single- and
double-spin asymmetries. This we do by studying the angular
distributions of the lepton pair in the vector boson rest-frame.

{\boldmath \section*{3. Large-$p_T$ Drell-Yan with $\gamma^*$--$Z^*$
interference}}
The subprocesses that contribute to the Drell-Yan process at
large $p_T$ are $q \bar q$ annihilation and $qg$ Compton scattering:
\begin{eqnarray}
\label{e7}
q + \bar q &\rightarrow \gamma^*,\ Z^* + g , \nonumber \\
q (\bar q) + g &\rightarrow \gamma^*,\ Z^* + q (\bar q),
\end{eqnarray}
followed by $\gamma^*,\ Z^* \rightarrow l^+l^-$. The Compton
subprocess dominates over the annihilation subprocess in
the case of $p$-$p$ scattering. Consequently, this process is
sensitive to gluon distributions in the $p$-$p$ case. We consider
polarised $p$-$p$ collisions (with the first proton labelled $A$
and the second labelled $B$). The corresponding partons
from protons $A$ and $B$ are labelled $a$ and $b$ and these carry
momentum fractions $x_1$ and $x_2$, respectively.
We will consider the case where both $A$ and $B$ are polarised
and also the case where $A$ is polarised and $B$ is not.

It is useful to think of the Drell-Yan process in terms of
a production process where the photon or the $Z$ is produced,
and a ``decay'' process for the boson into an $l^+ l^-$ pair.
We analyse the decay angular distributions of the lepton
in the rest frame of the $\gamma^*$ (or the $Z^*$) with
the $z$-axis taken to be the direction of the momentum
of the $\gamma^*$ or the $Z^*$ (in the c.m. frame). In
this frame, $\theta_l$ and $\phi_l$ are the polar and
azimuthal angles of the lepton.

We can factorise the Feynman amplitude for the Drell-Yan process $pp
\rightarrow l^+l^- X$ into two parts: one part of the amplitude
specifying the production of the virtual vector boson ($\gamma^*$
or $Z^*$), and the other describing the decay of this boson into a
pair of leptons. Thus, we may write (with $l^-$, $l^+$ standing
for the helicities of the lepton and the antilepton, respectively)
\begin{equation}
\label{e8}
M_{\lambda\mu;\rho\sigma}^{l^+l^-} = {1 \over M^2} \biggl \lbrack
M_{\lambda\mu;\rho\sigma}(\gamma) A_{\lambda}^{l^+l^-}(\gamma)
+ \chi(M) M_{\lambda\mu;\rho\sigma}(Z) A_{\lambda}^{l^+l^-}(Z)
\biggr \rbrack ,
\end{equation}
where the $M_{\lambda\mu;\rho\sigma}(\gamma, Z)$ are the production
amplitudes and $A_{\lambda}^{l^+l^-}(\gamma, Z)$ the decay
amplitudes for the $\gamma^*$ and the $Z^*$, and $\lambda, \mu,
\rho, \sigma$ are the helicities of the vector-meson, the
final parton and the initial partons $b$ and $a$, respectively;
$M$ is the invariant mass of the lepton pair and the function
$\chi(M)$ is given as
\begin{equation}
\label{e9}
\chi(M) = {M^2 \over M^2-M_Z^2+\mbox{ \rm i}M\Gamma_Z} ,
\end{equation}
where $M_Z$ and $\Gamma_Z$ are the mass and the width of the $Z$,
respectively. Squaring the amplitude in Eq.~(\ref{e8}) and summing
over all helicities except that of the vector boson, we arrive at
the following expression~:
\begin{eqnarray}
\label{e10}
\overline{ \vert M \vert^2}(\theta_l,\phi_l) =& \sum_{\lept}
{1 \over M^4}
\biggl \lbrack \rho_{\lambda\lambda^{\prime}}^{\gamma}
A^{l^+l^-}_{\lambda}(\gamma) A^{l^+l^-*}_{\lambda^{\prime}}(\gamma)
+ \vert \chi \vert^2 \rho_{\lambda\lambda^{\prime}}^{Z}
A^{l^+l^-}_{\lambda}(Z) A^{l^+l^-*}_{\lambda^{\prime}}(Z) \nonumber \\
&+\chi^* \rho_{\lambda\lambda^{\prime}}^{\gamma Z}
A^{l^+l^-}
_{\lambda}(\gamma) A^{l^+l^-*}_{\lambda^{\prime}}(Z)
+ \chi \rho_{\lambda\lambda^{\prime}}^{\gamma Z \dagger}
A^{l^+l^-}_{\lambda}(Z) A^{l^+l^-*}_{\lambda^{\prime}}(\gamma)
\biggr \rbrack ,
\end{eqnarray}
where the unnormalised density matrices $\rho_{\lambda\lambda^{\prime}}$
are defined as
\begin{equation}
\label{e11}
\rho_{\lambda\lambda^{\prime}}^{\gamma} = \sum_{\mu\rho\sigma}
f_{\sigma}^A(x_1)f_{\rho}^B(x_2)
M_{\lambda\mu;\rho\sigma}(\gamma)M^*_{\lambda^{\prime}\mu;\rho
\sigma}(\gamma) ,
\end{equation}
etc. The factors $f_{\sigma}^A(x_1),\ f_{\rho}^B(x_2)$ are the
appropriate density functions for partons of helicity $\rho$.
We write the Feynman amplitude for the decay as
\begin{equation}
\label{e12}
A^{l^+l^-}_{\lambda} = M_{\alpha} e^{i\phi_l\lambda} d^1
_{\lambda\alpha}(\theta_l) \hskip30pt \mbox{\rm (no sum on $\alpha$)},
\end{equation}
where $\alpha = l^- - l^+ = \pm 1$ only, because for fast leptons
only the helicities $(+\ -)$ or $(-\ +)$ are allowed. The
$M_{\alpha}$'s are essentially standard model coupling constants.
Then expanding the density matrices in terms of multipole parameters,
and integrating over $\phi_l$ yields
\begin{equation}
\label{e13}
\overline{ \vert M \vert^2}(\theta_l) = {\sqrt{\pi} \over M^4}
\sum_{l=0}^2 \biggl \lbrack
C_l^{\gamma}t_0^{l*}(\gamma)
+ \vert \chi \vert^2 C_l^{Z}t_0^{l*}(Z)
+2 \mbox{\rm Re} \lbrack \chi^* C_l^{\gamma Z}t_0^{l*}(\gamma Z)
\rbrack \biggr \rbrack Y_{l0}(\theta_l) ,
\end{equation}
where $Y_{l0}$ are the spherical harmonics and the $C_l$ are $decay\
parameters$ \cite{bls}, and we have
\begin{eqnarray}
\label{e14}
t_0^{l*} &=& \sum_\lambda \langle 1\lambda \vert 1\lambda ;
l0 \rangle \rho_{\lambda\lambda} , \nonumber \\
C_l &=& \sqrt{3} \sum_\alpha \langle l0 \vert 1\alpha ;
1 -\alpha \rangle M_{\alpha}^2 .
\end{eqnarray}
In the above equation $M_{\alpha}^2$ stands for $\vert M_{\alpha}
(\gamma) \vert^2$ and $\vert M_{\alpha}(Z) \vert^2$ for the
pure $\gamma$ and $Z$ terms and $M_{\alpha}(\gamma)M_{\alpha}^*(Z)$
for the interference terms.

Finally Eq.~(\ref{e13}) can be rewritten in the form
\begin{equation}
\label{e15}
\overline{ \vert M \vert^2}(\theta_l) = {4e^2 \sqrt{\pi} \over M^2}
\sum_{l=0}^2 D_l Y_{l0}(\theta_l) ,
\end{equation}
where the $D_l$'s are:
\begin{eqnarray}
\label{e16}
D_0 &=& t_0^0(\gamma) + (v^2+a^2)\vert \chi \vert^2 t_0^0(Z) + 2 v
(\mbox{\rm Re}\chi^*)t_0^0(\gamma Z), \nonumber \\
D_1 &=& -\sqrt{6} a \lbrack v \vert \chi \vert^2 t_0^1(Z) +
(\mbox{\rm Re}\chi^*)t_0^1(\gamma Z) \rbrack , \nonumber \\
D_2 &=& {1 \over \sqrt{2}} \lbrack t_0^2(\gamma) + (v^2+a^2)\vert
\chi \vert^2 t_0^2(Z) + 2 v (\mbox{\rm Re}\chi^*)t_0^2(\gamma Z)
\rbrack .
\end{eqnarray}
Here $v$ and $a$ are the vector and the axial couplings of the $Z$ to
the leptons. The expressions for the multipole parameters for the
annihilation and the Compton subprocesses, which involve the parton
distributions and the couplings, are presented in Appendix A.
The decay parameters $C_l$ are discussed in Appendix B.

\section*{4. Numerical results}
For the Drell-Yan process $AB \rightarrow \lept CX$,
($C$ is an associated quark- or gluon-initiated jet)
where the pair has transverse momentum $p_T$ and rapidity
$y_1$, the differential cross-section to produce the lepton at
angle $\theta_l$ (in the reference frame described in the previous
section) is then given as
\begin{equation}
\label{e17}
{d\sigma ^{AB \rightarrow \lept CX}
\over dp_T dy_1 d\tau d\mbox{\rm cos}\theta_l}=
\int dx_1 \sqrt{\pi} {\cal{F}} \sum_{l=0}^2 D_l Y_{l0}(\theta_l) ,
\end{equation}
where $\tau = M^2/s$, with $s$ the c.m. energy, and $x_1$
the momentum fraction of proton $A$ carried by parton
$a$. The factor $\cal{F}$ is given by
\begin{equation}
\label{e18}
{\cal{F}} = {\alpha x_T \over 48 \pi^2 \sqrt{s} M^2
 x_1 x_2 \biggl \lbrack x_1- {1\over 2}\sqrt{x_T^2+4\tau}e^{y_1}
\biggr \rbrack} ,
\end{equation}
where $x_T=2 p_T/\sqrt{s}$.

For each setting of the spins of the colliding protons the differential
cross-section in $\theta_l$ is controlled by the independent parameters
$D_{0,1,2}$, each of which contains information about the parton
distributions. To isolate information on the polarised distributions,
it is necessary to form asymmetries, either by reversing the spin
direction of one of the protons with the other being unpolarised
(single-spin asymmetries), or reversing the spin direction of one of
the protons, the other being polarised (double-spin asymmetries).

For each spin setting, the $D_j$'s can be projected out from a
knowledge of the angular distribution~:
\begin{equation}
\label{e19}
D_l =  {2\sqrt{\pi} \over {\cal{F}}}
\int {d\sigma ^{AB \rightarrow \lept CX}
\over dp_T dy_1 d\tau d\mbox{\rm cos}\theta_l}
Y_{l0}(\theta_l) \sin \theta_l d\theta_l ,
\end{equation}
Clearly $D_0$ is the simplest, being essentially the cross-section
to produce an $\lept$ pair with given $p_T$, $y_1$ and $\tau$.

The full set of asymmetries that one can, in principle, measure is the
following~:
\begin{eqnarray}
\label{e20}
A_s &=& {D_0^+ - D_0^- \over D_0^+ + D_0^-} =
{d\sigma_+ - d\sigma_-  \over d\sigma_+ + d\sigma_- }, \nonumber \\
A_d &=& {D_0^{++} - D_0^{-+} \over D_0^{++} + D_0^{-+}} =
{d\sigma_{++} - d\sigma_{-+}  \over d\sigma_{++} + d\sigma_{-+} }.
\end{eqnarray}
where the labels $+$, $-$ refer to the helicities of the protons $A$
and $B$, and for $j=1,2$,
\begin{eqnarray}
\label{e21}
A_s^j &=& {D_j^+ - D_j^- \over D_j^+ + D_j^-} , \nonumber \\
A_d^j &=& {D_j^{++} - D_j^{-+} \over D_j^{++} + D_j^{-+}}.
\end{eqnarray}
In what follows, we shall present results for all these asymmetries.
We shall assume that the jet recoiling from the lepton pair is not
detected and so we must integrate over $x_1$. We illustrate our results
for the case $y_1 =0$ and fixed values of $M$, and show how the
asymmetries vary with $p_T$. We take $\sqrt{s} = 500$~GeV to correspond
to RHIC energies.

In order to study the sensitivity of our asymmetries to the
polarised parton distributions we utilise the following range
of models~:

\begin{table}[htb]
\begin{center}
\begin{tabular}{ l  c  l }
Set I&:&$\Delta g$ large, $\Delta s = 0$,\\
Set II&:&$\Delta g$ and $\Delta s$ both moderately large,\\
Set III&:&$\Delta g = 0$, $\Delta s$ large,\\
\end{tabular}
\end{center}
\end{table}
\noindent which were described in Ref.~\cite{ours} and which all fit the
EMC data on $g_1^p(x)$. For the unpolarised distributions we use
Owens' Set~1.1 distributions \cite{owens}.

In Fig.~1, we show the calculated asymmetries for $M = $ 10 and
50~GeV and for $M=M_Z$ and with $\theta_l$ integrated over. The
figures in the top row are predictions for single-spin asymmetries,
the bottom row for double-spin asymmetries. In each figure, the three
curves correspond to the three different sets of polarised densities
used. We see that for small dilepton masses the single-spin asymmetries
are very small, but that they are appreciable for large dilepton
masses. This is because the single-spin asymmetry arises from the
parity-violating nature of the $Z$-fermion coupling, and hence is large
only when the effects of $Z$-interference are important. The double-spin
asymmetries are large over the whole range of dilepton masses.
In Fig.~1, there is a strong dependence upon the choice of
polarised parton distributions and the asymmetries, if measured
with reasonable accuracy, will be very helpful in teaching us
to what extent the various partons in a proton are polarised.
In Fig.~2 we have plotted the cross-section as a function of
$p_T$ for the three values of dilepton mass given above.
In Figs.~3 and 4 we have plotted the asymmetries $A_{s,d}^j\ (j=1,2)$.
Since the
structure of the contributions to $D_2$ is very similar to those of $D_0$
contribution, the corresponding asymmetries also look very similar
in shape. The single-spin asymmetries $A_s^{1,2}$ are large only
close to $M=M_Z$.

\section*{5. Conclusions}
Data on the polarised structure function $g_1^p(x)$ in deep-inelastic
scattering, from the EMC, SMC and SLAC experiments, if interpreted
in the naive parton model, suggest that only a small fraction of the
proton's spin arises from the spin of its quark constituents. One
explanation has the contribution of the valence quarks cancelled
by an unexpectedly large strange quark contribution. Another approach
argues that the naive parton model formulae are invalidated by the
axial anomaly and explains the data in terms of a large gluon
contribution to the proton's spin. Independent information on
the magnitude of the strange quark and gluon polarised distributions
is thus of great importance and needed urgently.
In this paper we have studied both the single- and the double-
spin asymmetries of the parameters which describe the angular
distribution of the lepton in the large transverse momentum
Drell-Yan process at RHIC energies ($\sqrt{s}=500$~GeV). At these
high energies, $Z^*$'s can contribute significantly to the Drell-Yan
process, inducing a parity-violating single-spin asymmetry. This
asymmetry may be measured in experiments where only one of the
initial particles is polarised.

We have discussed the framework in which we compute the angular
distributions of the leptons in the vector-boson rest-frame,
and have presented the complete results for the multipole parameters,
which describe the decay angular distributions of the lepton.
Using these angular distributions, we have constructed single- and
double-spin asymmetries, which we then studied numerically using
three sets of polarised parton distributions that are
consistent with the deep-inelastic scattering data.
We find that the single-spin asymmetries are measurably
large only at large dilepton masses, i.e. close to the $Z$ peak.
The double-spin asymmetries are large even at smaller dilepton
masses. Both the single-spin and the double-spin asymmetries are
sensitive to the polarised gluon distribution and may be used
as probes of the spin structure of the proton.

\clearpage
\section*{Appendix A}
In this appendix, we present the expressions for the density matrices
and the multipole parameters, for the annihilation and the Compton
processes.

The couplings of the $\gamma^*$ to both quarks and leptons is given by
the interaction term $-ieQ_f \gamma^{\mu}$, while the coupling of the
$Z^*$ to the fermions is given by the term $ ie \gamma^{\mu} (v_j - a_j
\gamma_5)$, where as usual
\begin{eqnarray}
v_j & = & {I_3^{j_L} - 2Q_j \sin^2 \theta_W \over 2 \sin \theta_W
\cos \theta_W} , \nonumber \\
a_j & = & {I_3^{j_L} \over 2 \sin \theta_W \cos \theta_W}
\end{eqnarray}
In the above equations, $I_3$ is the third component of the weak isospin,
$\theta_W$ the Weinberg angle and $Q_f,\ Q_j$ are the charges in
units of $e$. In what follows, we use $v$ and $a$
to denote the $Z^*$ couplings to leptons and $v_f$ and $a_f$ to quarks
of flavour $f$. The results are presented in terms of the variables
$\htau$ and $\hat \theta$, where $\hat \theta$ is the subprocess
scattering angle, and
\begin{equation}
\htau = {\tau \over x_1x_2} .
\end{equation}

The unnormalised density matrices for $Z^*$ production via the
annihilation process $q \bar q \rightarrow Z^* g$ are given as~:
\begin{eqnarray}
\label{ea1}
\rho_{11}(Z) \phantom{--}
  & = & K_{a} (1+\htau^2) \sum_f \biggl\lbrace \annia \biggl \lbrack
(v_f^2+a_f^2)\left (\cot^2 \t2 + \tan^2\t2\right ) \nonumber \\
&&+ 2v_fa_f\left (\cot^2 \t2 - \tan^2 \t2 \right ) \biggr \rbrack +\annib
\nonumber \\
&&\left \lbrack (v_f^2+a_f^2)\left (\cot^2 \t2 - \tan^2\t2 \right ) +
2v_fa_f \left(\cot^2 \t2 + \tan^2 \t2 \right ) \right\rbrack \biggr
\rbrace , \nonumber \\
\rho_{00}(Z) \phantom{--}
 & = & K_{a} 8\htau \sum_f \biggl\lbrace \annia (v_f^2+a_f^2)
 \nonumber \\
&&+ \annib (2v_fa_f) \biggr\rbrace , \nonumber \\
\rho_{1-1}(Z) \phantom{-}
 & = & K_{a} 4\htau \sum_f \biggl\lbrace \annia (v_f^2+a_f^2)
\nonumber \\
&&+ \annib (2v_fa_f) \biggr\rbrace , \nonumber \\
\rho_{10}(Z) \phantom{--}
 & = & K_{a} \sqrt{2\htau}(1+\htau) \sum_f \biggl\lbrace \annia
\biggl\lbrack (v_f^2+a_f^2)\left (\tan \t2 - \cot\t2\right ) \nonumber \\
&&- 2v_fa_f \left (\tan \t2 + \cot \t2\right )\biggr \rbrack -\annib
\times \nonumber \\
&&\left \lbrack (v_f^2+a_f^2)\left (\tan \t2 +
\cot\t2\right ) - 2v_fa_f \left(\tan \t2 - \cot \t2\right ) \right
\rbrack \biggr\rbrace ,
\nonumber \\
\rho_{0-1}(Z) \phantom{-}
 & = & K_{a} \sqrt{2\htau}(1+\htau) \sum_f \biggl\lbrace
\annia \biggl \lbrack (v_f^2+a_f^2)\left (\cot \t2 - \tan\t2\right )
\nonumber \\
&&- 2v_fa_f \left (\cot \t2 + \tan \t2\right ) \biggr \rbrack -\annib
\nonumber \\
&&\left \lbrack (v_f^2+a_f^2)\left (\tan \t2 + \cot\t2\right ) +
2v_fa_f \left(\tan \t2 - \cot \t2\right )\right \rbrack \biggr\rbrace ,
\nonumber \\
\rho_{-1-1}(Z)  & = & K_{a} (1+\htau^2) \sum_f \biggl\lbrace
\annia \biggl \lbrack (v_f^2+a_f^2)\left (\tan^2 \t2 + \cot^2\t2\right )
\nonumber \\
&&+ 2v_fa_f \left (\tan^2 \t2 - \cot^2 \t2\right ) \biggr \rbrack +\annib
\nonumber \\
&&\left \lbrack (v_f^2+a_f^2)\left (\tan^2 \t2 - \cot^2\t2\right )
+ 2v_fa_f \left (\tan^2 \t2 + \cot^2 \t2\right ) \right \rbrack
\biggr\rbrace ,
\end{eqnarray}
where the sum runs over all flavours. The overall factor $K_a$ is given
by
\begin{equation}
\label{ea2}
K_a = {4 \over 9} {e^2 g^2 \over (1-\htau)^2}.
\end{equation}
The unnormalised density matrices for $Z^*$
production via the Compton process $q g \rightarrow Z^* q$ are given as:
\begin{eqnarray}
\label{ea3}
\rho_{11}(Z) \phantom{--}
& = & K_{c} \sum_f \biggl\lbrace
\gqa \left \lbrack (v_f-a_f)^2\left ((1-\htau)^2 \sec^2 \t2 +
\htau^2 \cos^2 \t2
\tan^4 \t2 \right ) + (v_f+a_f)^2 \cos^2 \t2 \right \rbrack \nonumber \\
&&+\gqb \left \lbrack (v_f-a_f)^2\left ((1-\htau)^2 \sec^2 \t2
+ \htau^2 \cos^2 \t2
\tan^4 \t2 \right ) - (v_f+a_f)^2 \cos^2 \t2 \right \rbrack \nonumber \\
&&+\gqc \left \lbrack (v_f-a_f)^2\left ((1-\htau)^2 \sec^2 \t2
- \htau^2 \cos^2 \t2
\tan^4 \t2 \right ) + (v_f+a_f)^2 \cos^2 \t2 \right \rbrack \nonumber \\
&&+\gqd \left \lbrack (v_f-a_f)^2\left ((1-\htau)^2 \sec^2 \t2
- \htau^2 \cos^2 \t2
\tan^4 \t2 \right ) - (v_f+a_f)^2 \cos^2 \t2 \right \rbrack, \nonumber \\
\rho_{00}(Z) \phantom{--}
 & = & K_{c} 2\htau \sin^2 \t2 \sum_f \biggl\lbrace (\gqa-\gqd)
2(v_f^2+a_f^2) \nonumber \\
&&-4v_fa_f(\gqb-\gqc) \biggr\rbrace , \nonumber \\
\rho_{1-1}(Z) \phantom{-}
 & = & K_{c} \htau \sin^2 \t2 \sum_f \biggl\lbrace (-\gqa+\gqd)
2(v_f^2+a_f^2) \nonumber \\
&&+4v_fa_f(\gqb-\gqc) \biggr\rbrace , \nonumber \\
\rho_{10}(Z) \phantom{--}
 & = & K_{c} \sqrt{2\htau}\cos \t2 \sin \t2 \sum_f \biggl\lbrace
(-\gqa+\gqd) \biggl \lbrack (v_f-a_f)^2\htau \tan^2 \t2 + \nonumber \\
&&(v_f+a_f)^2\biggr \rbrack +(\gqc-\gqb) \left ((v_f-a_f)^2\htau
\tan^2 \t2 - (v_f+a_f)^2\right )
\biggr \rbrace , \nonumber \\
\rho_{0-1}(Z) \phantom{-}
 & = & K_{c} \sqrt{2\htau}\cos \t2 \sin \t2 \sum_f \biggl\lbrace
(\gqa-\gqd) \biggl ((v_f-a_f)^2 + (v_f+a_f)^2 \nonumber \\
&&\htau \tan^2 \t2\biggr )+(\gqb-\gqc) \left ((v_f-a_f)^2 - (v_f+a_f)^2
\htau \tan^2 \t2\right )
\biggr \rbrace , \nonumber \\
\rho_{-1-1}(Z)
& = & K_{c} \sum_f \biggl\lbrace
\gqa \left \lbrack (v_f+a_f)^2\left ((1-\htau)^2 \sec^2 \t2 +
\htau^2 \cos^2 \t2
\tan^4 \t2 \right ) + (v_f-a_f)^2 \cos^2 \t2 \right \rbrack \nonumber \\
&&+\gqb \left \lbrack -(v_f+a_f)^2\left ((1-\htau)^2 \sec^2 \t2
+ \htau^2 \cos^2 \t2
\tan^4 \t2 \right ) + (v_f-a_f)^2 \cos^2 \t2 \right \rbrack \nonumber \\
&&+\gqc \left \lbrack -(v_f+a_f)^2\left ((1-\htau)^2 \sec^2 \t2
- \htau^2 \cos^2 \t2
\tan^4 \t2 \right ) - (v_f-a_f)^2 \cos^2 \t2 \right \rbrack \nonumber \\
&&+\gqd \left \lbrack (v_f+a_f)^2\left ((1-\htau)^2 \sec^2 \t2
- \htau^2 \cos^2 \t2
\tan^4 \t2 \right ) - (v_f-a_f)^2 \cos^2 \t2 \right \rbrack, \nonumber \\
\end{eqnarray}
where, again the sum is over all flavours, and the overall factor
$K_c$ is given as
\begin{equation}
\label{ea4}
K_c = {1 \over 6} {e^2 g^2 \over (1-\htau)}.
\end{equation}
We can construct the unnormalised multipole parameters using the
expressions for the density matrices given above. For the annihilation
process, we obtain
\begin{eqnarray}
\label{ea5}
t_0^0(Z)
& = & 2 K_{a} \sum_f \biggl\lbrace
\annia (v_f^2+a_f^2) + \annib \nonumber \\
&&2v_fa_f \biggr\rbrace \left \lbrack (1+\htau^2) \left (\tan^2 \t2
+ \cot^2 \t2\right ) +4 \htau \right \rbrack , \nonumber \\
t_0^1(Z)
& = & \sqrt{2} K_{a} \sum_f \biggl\lbrace
\annia 2v_fa_f + \annib \nonumber \\
&&(v_f^2+a_f^2) \biggr\rbrace  \left \lbrack (1+\htau^2) \left (\cot^2
\t2 -  \tan^2 \t2\right ) \right \rbrack , \nonumber \\
t_0^2(Z)
& = & 2 \sqrt{1 \over 10} K_{a} \sum_f \biggl\lbrace
\annia (v_f^2+a_f^2) + \nonumber \\
&&\annib 2v_fa_f \biggr\rbrace \left \lbrack (1+\htau^2)
\left (\tan^2 \t2 + \cot^2 \t2\right ) -8 \htau \right \rbrack ,
\end{eqnarray}
and for the Compton process, we get
\begin{eqnarray}
\label{ea6}
t_0^0(Z)
& = & K_{c} \sum_f 2(v_f^2+a_f^2) \biggl\lbrace
\gqa \left \lbrack (1-\htau)^2 \sec^2 \t2 + \cos^2 \t2 \left (1 + \htau^2
\tan^4 \t2 \right )+2\htau \sin^2 \t2 \right \rbrack \nonumber \\
&&+\gqd \left \lbrack (1-\htau)^2 \sec^2 \t2 - \cos^2 \t2 \left
(1 + \htau^2
\tan^4 \t2 \right )-2\htau \sin^2 \t2 \right \rbrack \biggr\rbrace
\nonumber \\
&&-4v_fa_f \biggl\lbrace
\gqb \left \lbrack (1-\htau)^2 \sec^2 \t2 + \cos^2 \t2 \left (1 + \htau^2
\tan^4 \t2 \right )+2\htau \sin^2 \t2 \right \rbrack \nonumber \\
&&+\gqc \left \lbrack (1-\htau)^2 \sec^2 \t2 - \cos^2 \t2 \left
(1 + \htau^2
\tan^4 \t2 \right )-2\htau \sin^2 \t2 \right \rbrack \biggr\rbrace,
\nonumber \\
t_0^1(Z)
& = & {1 \over \sqrt{2}} K_{c} \sum_f -4v_fa_f \biggl\lbrace
\gqa \left \lbrack (1-\htau)^2 \sec^2 \t2 + \cos^2 \t2 \left (\htau^2
\tan^4 \t2 -1 \right ) \right \rbrack \nonumber \\
&&+\gqd \left \lbrack (1-\htau)^2 \sec^2 \t2 - \cos^2 \t2 \left (\htau^2
\tan^4 \t2 -1 \right ) \right \rbrack \biggr\rbrace \nonumber \\
&&+2(v_f^2+a_f^2) \biggl\lbrace
\gqb \left \lbrack (1-\htau)^2 \sec^2 \t2 + \cos^2 \t2 \left (\htau^2
\tan^4 \t2 -1 \right ) \right \rbrack \nonumber \\
&&+\gqc \left \lbrack (1-\htau)^2 \sec^2 \t2 - \cos^2 \t2 \left (\htau^2
\tan^4 \t2 -1 \right ) \right \rbrack \biggr\rbrace, \nonumber \\
t_0^2(Z)
& = & \sqrt{ 1\over 10} K_{c} \sum_f 2(v_f^2+a_f^2) \biggl\lbrace
\gqa \left \lbrack (1-\htau)^2 \sec^2 \t2 + \cos^2 \t2 \left (1 + \htau^2
\tan^4 \t2 \right )-4\htau \sin^2 \t2 \right \rbrack \nonumber \\
&&+\gqd \left \lbrack (1-\htau)^2 \sec^2 \t2 - \cos^2 \t2 \left
(1 + \htau^2
\tan^4 \t2 \right )+4\htau \sin^2 \t2 \right \rbrack \biggr\rbrace
\nonumber \\
&&-4v_fa_f \biggl\lbrace
\gqb \left \lbrack (1-\htau)^2 \sec^2 \t2 + \cos^2 \t2 \left (1 + \htau^2
\tan^4 \t2 \right )-4\htau \sin^2 \t2 \right \rbrack \nonumber \\
&&+\gqc \left \lbrack (1-\htau)^2 \sec^2 \t2 - \cos^2 \t2 \left
(1 + \htau^2
\tan^4 \t2 \right )+4\htau \sin^2 \t2 \right \rbrack \biggr\rbrace .
\end{eqnarray}
The multipole parameters given are only for quark-gluon Compton
scattering. The corresponding multipole parameters for
antiquark-gluon Compton scattering are given by
\begin{equation}
\label{ea7}
a_f \rightarrow - a_f
\end{equation}
in Eq. \ref{ea6}.

So far, we have listed only the $Z^*$ production density matrices
and multipole parameters. We need to have similar expressions for
$\gamma^*$ productions and the expressions for the interference
terms. These expressions can be easily obtained from the above
expressions for $Z^*$ by simple replacements of the couplings, which
we summarise in the following rules~:
\begin{enumerate}
\item
To obtain the $\gamma^*$ multipole parameters, the replacements
$a_f \rightarrow 0$, $v_f \rightarrow Q_f$, should be
made in the expressions for the $Z^*$ production multipole parameters.

\item
To obtain the multipole parameters corresponding to the interference
terms, the replacements $v_f^2 + a_f^2 \rightarrow -2Q_f v_f$, $2v_fa_f
\rightarrow -Q_f a_f$, should be made in the expressions for the
$Z^*$ production multipole parameters.
\end{enumerate}

Having listed all the multipole parameters, we are now in a position
to write down the complete expressions for the $D_j$'s defined in
Eq.~\ref{e16}. To present these in a compact form, we define the
following combinations of the couplings
\begin{eqnarray}
\label{ea8}
\alpha_f & = & Q_f^2 - 2vv_fQ_f \mbox{\rm Re} \chi^* + (v^2 + a^2)
(v_f^2+a_f^2) \vert \chi \vert^2  , \nonumber \\
\beta_f & = & 2a_fv_f \lbrack (v^2+a^2) \vert \chi \vert^2 -
vQ_f \mbox{\rm Re} \chi^* \rbrack , \nonumber \\
\gamma_f & = & aa_f \lbrack 2vv_f \vert \chi \vert^2 -
Q_f \mbox{\rm Re} \chi^* \rbrack , \nonumber \\
\delta_f & = &  v(v_f^2+a_f^2) \vert \chi \vert^2 -
v_fQ_f \mbox{\rm Re} \chi^* ,
\end{eqnarray}
and the functions,
\begin{eqnarray}
\label{ea9}
R_{\pm}(\htau, \hat \theta) & = & (1-\htau )^2 \sec^2 \t2 \pm
\left \lbrack
\cos \t2+ \htau \sin \t2 \tan \t2\right \rbrack^2 , \nonumber \\
S_{\pm}(\htau, \hat \theta) & = & (1-\htau )^2 \sec^2 \t2 \pm \cos^2 \t2
\left \lbrack \htau^2 \tan^4 \t2 - 1 \right \rbrack , \nonumber \\
T_{\pm}(\htau, \hat \theta) & = & R_{\pm}(\htau, \hat \theta)
\mp 6\htau \sin^2 \t2 , \nonumber \\
U(\htau, \hat \theta) & = & (1+\htau^2) \left \lbrack \tan^2 \t2 +
\cot^2 \t2
\right \rbrack +4 \htau , \nonumber \\
V(\htau, \hat \theta) & = & (1+\htau^2) \left \lbrack \tan^2 \t2
- \cot^2 \t2
\right \rbrack
\end{eqnarray}
We may now write each $D_j$ as
\begin{equation}
\label{ea10}
D_j = \sum_f \lbrack D_j^{q \bar q} + D_j^{qG} +D_j^{\bar q G} \rbrack
+ \lbrack \hat \theta \rightarrow \pi - \hat \theta \rbrack,
\end{equation}
where the sum is over flavours, and by $\lbrack \hat \theta \rightarrow
\pi - \hat \theta \rbrack$ is intended to imply the following
symmetrization
\begin{equation}
\label{ea11}
f_A(x_1)g_B(x_2)F(\hat \theta) \rightarrow g_A(x_1)f_B(x_2)
F(\pi - \hat \theta),
\end{equation}
where $f,g$ are parton densities pre-multiplying some function $F$ of
$\hat \theta$.

Then we have, for the annihilation subprocess,
\begin{eqnarray}
\label{ea12}
D_0^{q \bar q} & = & 2 K_a U(\htau, \hat \theta) \biggl\lbrace \annia
\alpha_f +\annib \beta_f \biggr \rbrace , \nonumber \\
D_1^{q \bar q} & = & 2 \sqrt{3} K_a V(\htau, \hat \theta) \biggl\lbrace
\annia \gamma_f +\annib \delta_f \biggr \rbrace , \nonumber \\
D_2^{q \bar q} & = & \sqrt{1 \over 5} K_a \lbrack U(\htau, \hat \theta)
-12 \htau \rbrack \biggl\lbrace \annia \alpha_f  \nonumber \\
&&\phantom{\sqrt{1 \over 5} K_a \lbrack U(\htau, \hat \theta)
-12 \htau \rbrack} + \annib \beta_f \biggr \rbrace ,
\end{eqnarray}
and for the Compton subprocess,
\begin{eqnarray}
\label{ea13}
D_0^{q G} & = & 2 K_c \biggl\lbrace \lbrack \gqa R_+(\htau, \hat \theta)
+\gqd R_-(\htau, \hat \theta) \rbrack \alpha_f \nonumber \\
&&- \lbrack
\gqb R_+(\htau, \hat \theta) +\gqc R_-(\htau, \hat \theta) \rbrack
\beta_f \biggr \rbrace , \nonumber \\
D_1^{q G} & = & 2 \sqrt{3} K_c \biggl\lbrace \lbrack \gqa S_+(\htau, \hat
\theta) +\gqd S_-(\htau, \hat \theta) \rbrack \gamma_f \nonumber \\
&&- \lbrack
\gqb S_+(\htau, \hat \theta) +\gqc S_-(\htau, \hat \theta) \rbrack
\delta_f \biggr \rbrace , \nonumber \\
D_2^{q G} & = & \sqrt{1 \over 5} K_c \biggl\lbrace \lbrack \gqa T_+(\htau,
\hat \theta) +\gqd T_-(\htau, \hat \theta) \rbrack \alpha_f \nonumber \\
&&- \lbrack
\gqb T_+(\htau, \hat \theta) +\gqc T_-(\htau, \hat \theta) \rbrack
\beta_f \biggr \rbrace .
\end{eqnarray}
$D_j^{\bar q G}$'s are obtained from $D_j^{q G}$'s by making the
replacements $\beta_f \rightarrow - \beta_f$ and $\gamma_f \rightarrow
- \gamma_f$.

\section*{Appendix B}
In this appendix, we discuss the decay process $V \rightarrow \lept$,
where $V$ denotes $\gamma^*$ or $Z^*$. We first list the Feynman
amplitudes $A_{\lambda}^{\lept}$, where $l^+$, $l^-$ refer to the
helicities of the antilepton and the lepton, respectively,
and $\lambda$ to the helicity of the vector-boson. Since the
vector-boson is massive (of mass $M$), we compute these amplitudes
in the vector-boson rest frame. In this frame, we take the polar
and azimuthal angles of the lepton to be $\theta_l$ and $\phi_l$,
respectively.

For fast leptons, because of the vector and the axial-vector couplings,
only the helicities (+,-) or (-,+) can occur and the Feynman amplitude
has the form
\begin{equation}
\label{eb1}
A_{\lambda}^{\lept} = M_{\alpha} e ^{i\phi_l \lambda} \mbox{\rm d}^1
_{\lambda \alpha} (\theta_l)
\end{equation}
where $\alpha = l^- - l^+ = \pm 1$.

Upon calculating the Feynman amplitudes, we find that
\begin{equation}
\label{eb2}
M_{\alpha} = \sqrt{2} i M \beta(\alpha )
\end{equation}
where $\beta(\alpha )$ depends only on the electroweak coupling
constants. For the decay of the photon one finds
\begin{equation}
\label{eb3}
\beta^{(\gamma)}(1) = ie
\end{equation}
and, by parity invariance, we get
\begin{equation}
\label{eb4}
\beta^{(\gamma)}(-1) = \beta^{(\gamma)}(1)
\end{equation}

Then, from Eq.~\ref{e14}, the decay parameters $C_l$ are given by
\begin{equation}
\label{eb9}
C_l^{(\gamma)}(00;00) =  \sqrt{3} (2M^2) \sum_{\alpha} \vert
\beta^{(\gamma)}(\alpha) \vert^2 \langle l0 \vert 1 \alpha ; 1 -
\alpha \rangle ;
\end{equation}
which gives
\begin{eqnarray}
\label{eb10}
C_0^{(\gamma)}(00;00) & = & 4e^2M^2 , \nonumber \\
C_1^{(\gamma)}(00;00) & = & 0 , \nonumber \\
C_2^{(\gamma)}(00;00) & = & 2\sqrt{2}e^2M^2 .
\end{eqnarray}
For the $Z^*$, $\beta^{(Z)}(1) = ie  (v-a)$ and $\beta^{(Z)}(-1)
= ie (v+a)$, and so we get
\begin{eqnarray}
\label{eb11}
C_0^{(Z)}(00;00) & = & (v^2+a^2) C_0^{(\gamma)}(00;00) , \nonumber \\
C_1^{(Z)}(00;00) & = & -4\sqrt{6} M^2 e^2 va , \nonumber \\
C_2^{(Z)}(00;00) & = & (v^2+a^2) C_2^{(\gamma)}(00;00) .
\end{eqnarray}
Finally, for the interference term, we obtain
\begin{eqnarray}
\label{eb12}
C_0^{(\gamma Z)}(00;00) & = & v C_0^{(\gamma)}(00;00) , \nonumber \\
C_1^{(\gamma Z)}(00;00) & = & -2\sqrt{6} M^2 e^2 a , \nonumber \\
C_2^{(\gamma Z)}(00;00) & = & v C_2^{(\gamma)}(00;00) .
\end{eqnarray}

\newpage
\section*{Figure captions}
\renewcommand{\labelenumi}{Fig. \arabic{enumi}}
\begin{enumerate}
\item   
Asymmetries for the Drell-Yan process at $\sqrt{s}=500$~GeV, as a
function of $p_T$. The asymmetries shown in this figure are
constructed out of the cross sections differential in $p_T$, $y_1$,
and $\htau$ (the jet rapidities have been integrated over), with
$\htau$ fixed at $M/\sqrt{s}$ and $y_1=0$.
The cross sections are integrated over the lepton angle $\theta_l$.
The dilepton mass $M$ is shown on top of the
figures. The full, dashed and dotted lines correspond respectively
to Set~I, II and III polarised densities, described in the text.
The upper row of plots are for the single-spin asymmetries $A_s$, the
lower for the double-spin asymmetries $A_d$. Note that the scale for the
first two plots is much smaller than for the rest.
\item   
Cross-sections for the Drell-Yan process at $\sqrt{s}=500$~GeV, as a
function of $p_T$. The caption $d\sigma/d\Gamma$ refers to
$d\sigma/dp_T dy_1 d\htau $,
with $\htau$ fixed at $M/\sqrt{s}$ and $y_1=0$.
The cross sections are integrated over the lepton angle $\theta_l$.
The full, dashed and dotted lines correspond respectively
to $M$=10~GeV, $M$=50~GeV and $M=M_Z$, where $M$ is the dilepton mass.
The cross sections are in nb/GeV.
\item   
Same as in Fig.~1, but for the asymmetries $A_s^1$ and $A_d^1$
defined in Eq.~\ref{e21}.
\item   
Same as in Fig.~1, but for the asymmetries $A_s^2$ and $A_d^2$
defined in Eq.~\ref{e21}.
\end{enumerate}
\end{document}